\documentstyle[prl,aps,psfig]{revtex}
\def \bp {{\bf p}}
\def \bg {{\bf g}}
\def \bR {{\bf R}}
\begin{document}


\widetext
\title{ A microscopic approach to phase transitions in quantum systems}
\author { Pietro Gianinetti$^{1,2}$ and Alberto Parola$^{1,3}$
\footnote{Fax: +39-02-2392482, e-mail: parola@mi.infm.it} }
\address{ 
$^1$ Istituto Nazionale per la Fisica della Materia \\
$^{2}$ Dipartimento di Fisica, Universit\'a di Milano, Via Celoria 16,
Milano, Italy \\
$^3$ Dipartimento di Scienze Fisiche, Universit\'a dell'Insubria, Via Lucini 3
Como, Italy } 

\maketitle
\begin{abstract}
We present a new theoretical approach for the study of 
the phase diagram of interacting quantum particles: bosons, fermions or spins.
In the neighborhood of a phase transition, 
the expected renormalization group structure is recovered
both near the upper and lower critical dimension. 
Information on the microscopic hamiltonian is also retained and
no mapping to effective field theories is needed.
A simple approximation to our formally exact equations is
studied for the spin-$S$ Heisenberg model in three dimensions
where explicit results for critical exponents, critical temperature and 
coexistence curve are obtained. 
\end{abstract}

Several physical systems, ranging from magnets to superfluids and 
superconductors,
display rich phase diagrams in a temperature regime where quantum effects
cannot be neglected. Different scenarios, characterized by competing
order parameters and zero temperature phase transitions have been 
recently advocated also in the framework of high temperature superconductivity
where antiferromagnetic order, Cooper pairing and, possibly, phase separation 
are at play in the same region of the phase diagram \cite{htc}.
A satisfactory understanding of phase transitions in
quantum models has been attained years ago through the seminal work
by Hertz \cite{quantum} who showed that, at low energy and long 
wavelengths, quantum models may be described by a suitable classical
action. However, a quantitative theory
of the thermodynamic behavior is still lacking and we mostly rely on 
mean field approaches or weak coupling renormalization group
(RG) calculations \cite{mfrg}, applied to quantum systems via the mapping to 
the appropriate effective
field theory. In particular, the interplay between thermal and quantum
fluctuations is expected to give rise to crossover phenomena
whose extent strongly depends on the microscopic features of the system.
Even for the most extensively studied models, like the Heisenberg
antiferromagnet, our knowledge of the phase diagram is in fact limited, 
and the first precise finite temperature simulation attempting to fill this gap
has become available only recently \cite{sandvik}.
By contrast, in classical models, numerical simulations are quite efficient
even in the neighborhood of critical points \cite{mc} and,
from the analytical side, microscopic approaches especially devised for
the quantitative description of the phase diagram of classical fluids
and magnets are available. For instance, the hierarchical reference
theory of fluids (HRT) \cite{hrt} has proven quite accurate in
locating the phase transition lines both in lattice and in continuous
models.  

In this Letter we sketch the derivation of the quantum hierarchical 
reference theory of fluids (QHRT) which we then apply to the 
Heisenberg antiferromagnet.  We will
demonstrate that the known renormalization group equations near four and
near two dimensions are naturally recovered within our approach, which
therefore unifies two complimentary techniques. On approaching the
critical point, the spin velocity vanishes according to the
expected dynamical critical exponent for an antiferromagnet. 
Finally, the phase diagram of this model in three dimensions 
is computed by numerical integration of a simple approximation to the
the QHRT equations, providing a concrete application of our general approach.

The starting point is a microscopic, many body hamiltonian $H$ written
as the sum of a reference part $H_0$ and an interaction term $V$. 
The interaction is assumed to be bilinear in some operator $\rho(r)$,
which is assumed either linear in bosonic operators or quadratic in fermionic
ones:
\begin{equation}
V={1\over 2}\int dx dy \rho(x) w(x-y) \rho(y) 
\end{equation}
with a non singular (i.e. Fourier transformable) two body potential $w$.
The properties of the reference system under the action of
an external field $h$ coupled to the order parameter $\rho(r)$ are 
supposed known. No specific  assumption on the reference system is
made: in particular we do not need that $H_0$ corresponds to
a non interacting system, where Wick theorem applies (such a feature is
crucial in setting up the QHRT equations).
These requirements are indeed rather general and include several models 
of current interest in many body physics: quantum magnets (where
$\rho(r)$ represents the local spin variable), fermionic systems, like the
Hubbard model, or even the Holstein model for the
electron-phonon problem.

The first task is to build up a formal perturbative
expansion of the partition function of the model: $Z={\rm Tr}\exp(-\beta H)$.
Following a standard procedure \cite{fetter}, $Z/Z_0$ can be written as 
the average over the reference distribution function of an imaginary 
time evolution operator $U(\beta)$. When this operator is 
written as a power series  of 
the interaction $w(r)$, we formally recover a perturbative
expansion identical to that of a classical partition function for a 
$(d+1)$ dimensional model. The additional ``temporal" dimension is 
limited to the interval $(0,\beta)$ and $w(r)\delta(t)/\beta$ plays 
the role of classical two body interaction $w_c(r,t)$.
The reference system of the associated classical
model is implicitly defined by requiring that its 
correlation functions coincide with those of the quantum reference 
hamiltonian $H_0$. 
{\sl Approximate} mappings 
between a quantum model and an effective classical system have 
been proposed and studied in the literature \cite{tognetti}
in order to clarify the role of thermal and quantum 
fluctuations. The novelty of our approach is that $i)$ it is {\sl exact} 
and $ii)$ it applies to all temperature regimes, including the $T\to 0 $
limit. Having reduced the quantum problem to a
classical one, we can directly apply the techniques developed in that 
framework.  In particular, the already mentioned HRT is an 
implementation of the momentum space renormalization method 
which preserves information on the details of the microscopic hamiltonian. 
In HRT, different Fourier components of the two body interaction
are included gradually, starting from the shortest wavelength:
physically this corresponds to a smooth turning on of fluctuations over
larger and larger lengthscales. This procedure can be carried out exactly 
by defining a sequence of auxiliary systems interacting via
a potential whose Fourier components coincide with $w(k)$ for
$k>Q$ and vanish elsewhere. As a result, we obtain
a set of coupled differential equations
expressing the change in the free energy and in the correlation
functions of the model when a given Fourier component $Q$ of the
potential is included. The initial condition represents the system
in which fluctuations are frozen, and in fact coincides with
the known mean field result. As the {\sl cut-off} wavevector
$Q\to 0$, the fully interacting system
is recovered. Mean field {\sl approximation} therefore
corresponds to neglecting the change in the properties of the
model as described by our differential equations. Details can be
found in Ref. \cite{hrt}.

As an example, we study the the spin-$S$ antiferromagnetic 
Heisenberg model on a hypercubic lattice in $d$ dimension:
\begin{equation}
H=H_0+V=h\sum_\bR e^{i\bg\cdot\bR} S^z_\bR+
J \sum_{<\bR,\bR^\prime>} {\bf S_R}\cdot {\bf S}_{{\bf R}^\prime}
\end{equation}
where the sum is restricted to nearest neighbors and ${\bf g}$ is
the antiferromagnetic wavevector of components $g_i=\pi$.
In applying the HRT approach to a quantum model, we decided to
impose the cut-off $Q$ only to the spatial Fourier
components of $w_c(r,t)$. Different physical models might require
other choices of the cut-off, whose only role is  to
continuously connect the reference system
to the fully interacting one, which is recovered in the $Q\to 0$ limit.
The exact evolution equation for the Helmholtz free energy 
density $a$ of the system describes how $a$ is modified due to 
a change in the cut-off $Q$:
\begin{eqnarray}
\frac{d \, a^Q}{dQ}=\frac{1 }{2\beta} 
\int_{\bp\in\Sigma_Q} \sum_{\omega}
&\Big \{&  2\,\ln \left [ (1-F^Q_{xx}(\bp,\omega)w(\bp))(1+F^Q_{xx}
(\bp^{\prime},\omega)w(\bp))
-F_{xy}^Q(\bp,\omega)F_{xy}^Q(\bp^{\prime},\omega) w(\bp)^2 \right ]\nonumber\\
&+& \ln\left [(1-F^Q_{zz}(\bp,\omega)w(\bp))(1+F^Q_{zz}(\bp^{\prime},\omega)
w(\bp))\right]\Big \} 
\label{qhrt}
\end{eqnarray}
Here $\bp^{\prime}=\bp+\bg$, 
$w(\bp)=2J\gamma(\bp)=2J\sum_i \cos(p_i)$ is the Fourier transform of the
interaction, the summation is over the Matsubara frequencies
$\omega_n=2\pi n/\beta$, the $(d-1)$ dimensional integral 
is restricted to the
surface $\Sigma_Q$ defined by $\gamma(\bp)=-\sqrt{d^2-Q^2}$ 
and the functions $F^Q_{ij}(\bp,\omega)= <S^i(\bp,\omega)
S^j(-\bp,-\omega)>$ 
are the Fourier transforms of the spin-spin dynamical correlation functions
(in imaginary time) for a system where only fluctuations of
wavevector $\bp$ such that $|\gamma(\bp)|<\sqrt{d^2-Q^2}$ are included.
The isotropy of the model implies $F_{yy}=F_{xx}$ and $F_{yx}=-F_{xy}$.
Analogous equations can be derived for the many spin correlation
functions of the model. The (infinite) set of differential equations
forms the QHRT hierarchy.
When $Q=d$, fluctuations are neglected and the exact initial condition
for the first QHRT equation (\ref{qhrt}) coincides with the mean field
free energy density. The magnetic structure factors at $Q=d$ 
can be explicitly written as:
\begin{eqnarray}
F^Q_{xx}(\bp,\omega)&=&{\mu_\perp-w(\bp)\over m^{-2}\omega^2+
\mu_\perp^2-w(\bp)^2};\nonumber \\
F^Q_{xy}(\bp,\omega)&=&{m^{-1}\omega\over
m^{-2}\omega^2+\mu_\perp^2-w(\bp)^2};\nonumber \\
F^Q_{zz}(\bp,\omega)&=&{\delta_{\omega,0}\over\mu_{||}+w(\bp)}
\label{fij}
\end{eqnarray}
where $m$ is the staggered magnetization and $\mu_\perp,\,\,
\mu_{||}$ are known functions of $m$. For $S=1/2$: 
$\mu_\perp=2T\,m^{-1}\tanh^{-1}(2m)$ and $\mu_{||}=4T(1-4m^2)^{-1}$. 
Note that the dependence of
the transverse magnetic structure factors in (\ref{fij}) on 
frequency and momentum is consistent with the 
first order spin wave result at zero temperature and reproduces the
known single mode approximation which well represents
antiferromagnetic correlations at low temperatures
\cite{stringari,singh}. The longitudinal
correlations in equation (\ref{fij}) are instead purely classical
and, as such, satisfy the relationship $T\,\chi_{||}(k)=S_{||}(k)$
between longitudinal susceptibility $\chi_{||}(k)$ and the corresponding 
static structure factor $S_{||}(k)$. 

Equation (\ref{qhrt}), although
formally exact, is not closed because the evolution of the free energy
depends on the unknown magnetic structure factors of the model 
$F_{ij}^Q(\bp,\omega)$. Therefore, we have to introduce some approximate
parametrization of the structure factors in terms of the free energy.
The simple approximation we have studied is to retain the form 
of $F_{ij}^Q(\bp,\omega)$ as given in Eq. (\ref{fij}) but imposing 
thermodynamic
consistency in order to determine the two scalar parameters $\mu_\perp$
and $\mu_{||}$. More precisely, for every $Q$ we related the transverse
and longitudinal staggered susceptibilities to the free energy 
via the exact sum rules: 
\begin{eqnarray}
(\mu_\perp-2dJ)^{-1}&=&F_{xx}^Q(\bg,\omega=0)=
m(\partial a^Q/\partial m)^{-1} \nonumber \\
(\mu_{||}-2dJ)^{-1}&=&F_{zz}^Q(\bg,\omega=0)=
(\partial^2 a^Q/\partial m^2 )^{-1} 
\label{sum}
\end{eqnarray}
From the adopted structure of the dynamical correlation functions,
we also obtain the relationship between the parameters entering
$F_{ij}^Q$ and the zero temperature non linear sigma model coupling constants:
uniform transverse susceptibility $\chi_0=1/(4d)$, spin wave velocity
$c=\sqrt{4d}\,m$ and spin stiffness $\rho_s=m^2$. The hydrodynamic 
relation $\chi_0 c^2=\rho_s$ is automatically satisfied by our ansatz
for arbitrary spontaneous magnetization $m$.
It is interesting to note that from our parametrization,
the scaling of the spin wave velocity $c$ on approaching the 
critical temperature ($t=(T_c-T)/T_c\to 0$)
gives $c\propto m$ that is $c\propto t^\beta$ along the coexistence curve. 
The dynamic scaling hypothesis instead predicts $c\propto t^{\nu(z-1)}$
where $z$ is the dynamical critical exponent. By use of scaling laws
and recalling that in our approximation the correlation critical
exponent vanishes ($\eta=0$), we get $z=d/2$ which is the expected
result for an antiferromagnet (i.e. model G) \cite{hohenberg}.
Equation (\ref{qhrt}), together with (\ref{fij}) and (\ref{sum}) 
give rise to a partial differential equation for the free energy 
density of the Heisenberg model $a^Q(m)$ as a function of the 
cut-off $Q$ and of the
magnetization $m$. The frequency sum can be carried out analytically
giving the final equation:
\begin{equation}
\frac{d \, a^Q}{dQ}=\frac{1 }{2\beta} 
\int_{\bp\in\Sigma_Q} 
\left \{ 4\,\ln \left [{\sinh\left ({1\over 2}\beta m\mu_\perp\right )\over 
\sinh\left ({1\over 2}\beta m\sqrt{\mu_\perp^2-w(\bp)^2}\right )}\right ]
+ \ln\left [{\mu_{||}^2\over\mu_{||}^2-w(\bp)^2} \right]\right \} 
\label{qhrt2}
\end{equation}
We numerically solved this partial differential equation for several 
values of the
spin $S$ and different temperatures in order to study the phase diagram of
this system. Note that $S$ just enters the theory through the 
initial condition $a^Q(m)$ at $Q=d$, while the form of the differential 
equation is unaffected by $S$. Before showing the numerical results, however, 
it is useful to discuss the behavior of Eq. (\ref{qhrt2}) near
a phase transition. In particular, we studied the neighborhood of the
critical point (in $d>2$) and the low temperature region. 
Both at the critical point and along the coexistence curve the 
susceptibilities diverge due to the presence of critical fluctuations
and Goldstone bosons, respectively. From Eq. (\ref{sum}) we conclude that 
at long wavelengths (i.e. $Q\to 0$) and near a phase
transition we have $\mu_\perp\sim\mu{||}\sim 2dJ$. In this region
equation (\ref{qhrt2}) simplifies and, by rescaling the free energy
as $a^Q Q^{-d}$ and the magnetization as $m Q^{(2-d)/2}$, it reduces
to the RG equation obtained by Stanley {\it et al.}
\cite{stanley} for a $O(3)$ symmetric $\phi^4$ hamiltonian. Such an
equation has been analyzed near four dimension and proved to give
the correct critical exponents to first order in the $\epsilon=4-d$
expansion. In three dimensions, the numerical solution of the 
universal fixed point equation gives for the correlation length
critical exponent the result $\nu=0.826$, to be compared with the
accepted value $\nu=0.71$. The other critical exponents follow from
the scaling laws, noting that our analytical form of the two point
functions (\ref{fij}) forces the anomalous dimension exponent to vanish
$\eta=0$. We therefore find non classical exponents in three dimensions.
Special care must be paid when dealing with the $T\to 0 $ limit of
our equation. In this case, the asymptotic form of the equation changes
and it can be shown to give rise, near a hypothetical quantum critical
point, to critical exponents appropriate for
a $O(3)$ model in $d+1$ dimensions as expected. 
A separate analysis should be carried out in the low temperature phase.
If symmetry is spontaneously broken, following Chakravarty {\it et al.}
\cite{chn}, we may ask how quantum and thermal
fluctuations modify the zero field magnetization. In order to 
answer this question we perform a Legendre transform on 
our equation (\ref{qhrt2}): we first derive it with respect to the 
magnetization $m$ obtaining an evolution equation for the magnetic
field $h^Q(m)$ at fixed $m$. 
Then we find the equation governing the evolution of 
the spontaneous magnetization $m^Q$ implicitly defined by the 
requirement $h^Q(m^Q)=0$ 
at every $Q$. This procedure gives rise to a differential
equation for $m^Q$. In the $Q\to 0$ limit, taking into account that 
the longitudinal susceptibility diverges more slowly than $Q^{-2}$, 
QHRT reduces to a simple ordinary differential equation:
\begin{equation}
{d m^Q\over dQ}=K_d \left({Q \over \sqrt{d}}\right)^{d-2} 
\left [ \tanh \left (Q \beta \, m^Q\right)\right ]^{-1}
\label{mag}
\end{equation}
where $K_d$ is a geometrical factor (ratio between the solid angle and
volume of the Brillouin zone). By introducing the rescaled variable
$g=\sqrt{4d}(Q/\sqrt{d})^{d-1}/m^Q$, equation (\ref{mag}) becomes identical,
to order $g^2$, to the known weak coupling RG
equations for the non linear sigma model applied
by Chakravarty {\it et al.} \cite{chn} to the analysis of the 
antiferromagnetic Heisenberg model at long wavelengths and low
temperatures. As an example, we plot in Fig. 1 the RG flux of $g^Q$
obtained by the integration of the {\sl full} QHRT equation (\ref{qhrt2}).
We clearly see the effect of the unstable zero temperature weak
coupling fixed point while, for the nearest neighbor Heisenberg model, 
the other fixed point ($g_c$), governing the quantum critical regime, has no
effect on the RG trajectories.
This analysis shows that the single mode approximation to 
QHRT reproduces the correct long wavelength structure both near four and two
dimensions. Furthermore, we expect QHRT to be superior to the weak coupling
renormalization group equations because our non perturbative approach 
also describes the critical region and the high temperature regime
where $m^Q\to 0$ as $Q\to 0$, corresponding to $g\to \infty$ i.e. to
the strong coupling phase of the non linear sigma model.

Finally, we present few results of the numerical integration of Eq.
(\ref{qhrt2}) in three dimensions. As already pointed out in the
classical case \cite{hrt}, the HRT approach is able to correctly
implement Maxwell construction at first order phase transitions
and in fact the free energy density
$a^Q(m)$ at the end of the integration, i.e. in the $Q\to 0$ limit,
becomes rigorously flat in a finite region of the magnetization
axis for $T<T_c$. Therefore it is easy to extract from the numerical output 
the critical temperature and the coexistence curve, shown 
in Fig. 2 for several values of the spin $S$. The zero temperature 
limits of this curve agree within about $2\%$ with the accepted 
estimates based on spin wave theory, Monte Carlo simulations or 
series expansions. Regrettably, for the spontaneous magnetization 
at finite temperature there are just few available results going
beyond mean field approaches.  Simulation data for the classical
$S\to\infty$ case \cite{binder} and recent series expansion for the $S=1/2$
model \cite{kok} seem to give somewhat larger coexistence regions. 
However, we believe that a more systematic analysis of these models by 
accurate numerical techniques is necessary before reaching a definite
conclusion on the accuracy in the determination of the coexistence
curve. The critical temperature for the classical model is 
known by several methods \cite{mc,domb} to be $T_c=1.443 \,J$ while
for the $S=1/2$ case it has been recently estimated as $T_c=0.946 \,J$
\cite{sandvik} by use of a newly developed Quantum Monte Carlo method
and $T_c=0.93 \,J$ \cite{kok} by series expansions.
Our results are a few percent lower, being $T_c=1.419 \,J$ for
$S=\infty$ and $T_c=0.90 \,J$ for $S=1/2$. From the solution of the QHRT
equation we also obtain other important information on the model, for 
instance, the equation of state, the specific heat and also the temperature
dependent dynamical structure factors, via analytic continuation of 
the adopted expressions (\ref{fij}). In order to improve 
the QHRT results we have just discussed, other approximate 
expressions for the magnetic structure factors should be 
examined, possibly keeping the same form (\ref{fij}) but allowing for a
non trivial renormalization factor for the uniform susceptibility.
This method can be applied in a straightforward way to other models
of interest in quantum many body physics, like the Hubbard model,
and may help to determine the location of the magnetic phase transitions
and the possible occurrence of phase separation in a purely repulsive 
electron system.

\begin{figure}[htbp]
\protect
\centerline{\psfig{figure=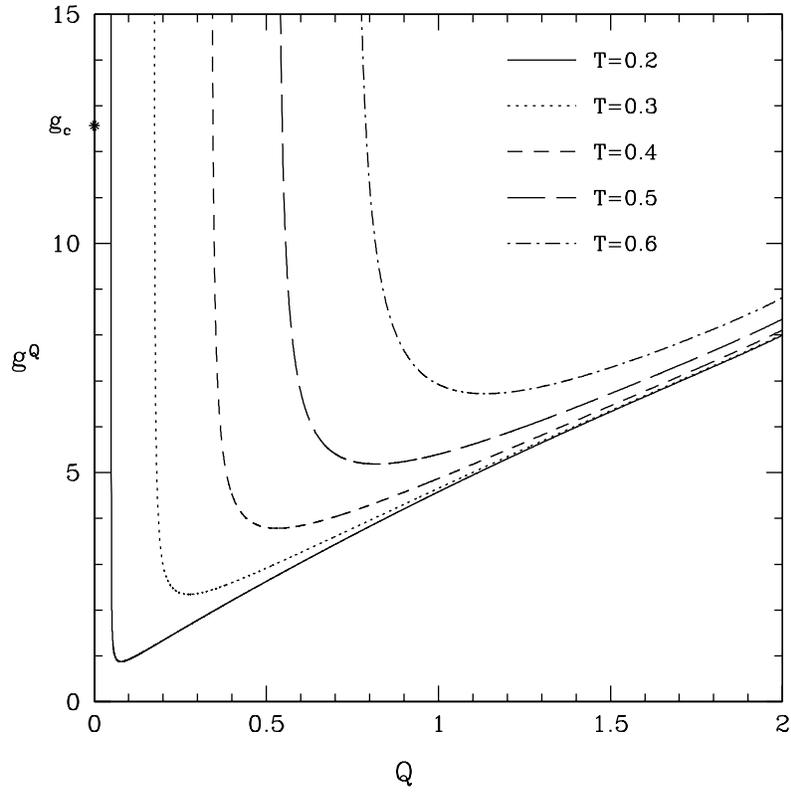,width=15cm}}
    \caption{RG trajectories for the two dimensional Heisenberg model
computed via numerical integration of the QHRT equation.}
\end{figure}

\newpage

\begin{figure}[htbp]
\protect
\centerline{\psfig{figure=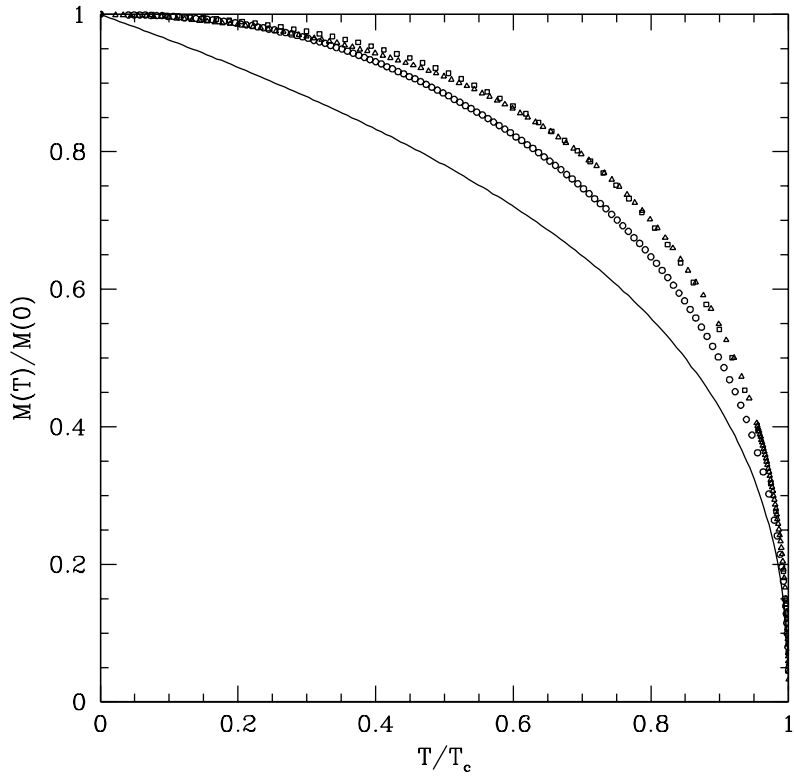,width=15cm}}
    \caption{Reduced spontaneous magnetization as a function of temperature
for different values of the spin: $S=1/2$ (triangles) $S=1$ (squares)
$S=5/2$ (circles) and $S=\infty$ (full line).}
\end{figure}
\end{document}